\documentclass[print, 
               notlongauthorlist 
              ]{nsr}

\usepackage{rotating}

\volume{00}

\artnum{00}

\firstpage{1}

\datesubmitted{XX Month 2026}

\doinum{doi/number}

\copyrightyear{2026}

\author[1,2]{S. V. Troitsky\footnote{Corresponding author e-mail address: \tt st@inr.ac.ru}$^{\orcidlink{0000-0001-6917-6600}}$}

\affil[1]{Institute for Nuclear Research,
of the Russian Academy of Sciences, 
60th October Anniversary Prospect 7a, Moscow 117312, Russia}
\affil[2]{Physics Department, Lomonosov Moscow State University, 1-2 Leninskie Gory, Moscow 119991, Russia}

\runauth{S. V. Troitsky}

\title{Combined constraints on the diffuse flux of cosmic neutrinos between $10^{16}$~eV and $10^{26}$~eV}

\begin{document}
\maketitle

\begin{abstract}
\noindent
Experimental constraints on the diffuse neutrino flux above the PeV range have been obtained with Cherenkov neutrino telescopes, air-shower arrays, radio detectors in and above polar ice, and radio observations of the Moon. Published results use different flavor, energy-bin, and statistical conventions, so their flux limits cannot be combined directly. We reconstruct energy-dependent exposures of the published searches and place them in a common convention for the total all-flavor $\nu+\bar\nu$ flux over $10^{16}$--$10^{26}$~eV. Published event counts and expected backgrounds are combined with a Poisson likelihood, and 90\% C.L. quasi-differential limits are obtained for one-decade $E_\nu^{-1}$ test spectra using a one-sided profile-likelihood construction. Folding theoretical spectra of cosmogenic neutrinos and selected new-physics scenarios with the combined energy-dependent exposure yields constraints on their flux normalizations. This homogeneous analysis provides a reproducible observational benchmark across ten decades in neutrino energy and a common reference for current and projected searches.
\end{abstract}

\begin{keyword}
neutrino astronomy, ultra-high-energy cosmic rays
\end{keyword}

\section{Introduction}
\label{sec:intro}

High-energy extraterrestrial neutrinos, detected by IceCube \cite{IceCubeDiscovery2013,IceCube-Astro2022} and Baikal-GVD \cite{BaikalGVDDiscovery2023,Baikal-diffuse2}, trace the most energetic non-thermal processes in the Universe. They propagate essentially undeflected and unattenuated from their production sites, bringing multimessenger information about interactions of relativistic hadrons. Above the PeV scale, neutrino observations become particularly important for the study of ultra-high-energy cosmic rays (UHECRs): neutrinos may be produced either inside or near their accelerators or during the propagation of UHECRs through cosmic radiation backgrounds.  The attenuation of UHECRs on cosmic radiation backgrounds was identified by Greisen \cite{Greisen1966} and by Zatsepin and Kuzmin \cite{ZatsepinKuzmin1966}, while the associated ultrahigh-energy neutrino component was pointed out soon afterwards by Berezinsky and Zatsepin \cite{Berezinsky:cosmogenic}.  These cosmogenic neutrinos provide a probe of the composition, maximal energies, cosmic-ray spectra and source cosmological evolution of UHECR.  Present UHECR spectrum and composition data allows for different source scenarios \cite{AugerUHECR2023,BergmanICRC2021}, with correspondingly different cosmogenic-neutrino expectations \cite{KuznetsovCosmogenic2026,AlhebsiCosmogenic2026}.  At still higher energies, UHE-particle production may also arise from superheavy-particle decay \cite{KuzminTkachev1998}, which can be tested directly with UHE-neutrino data \cite{KM3NeTSHDM2026}, or from topological defects \cite{BerezinskyVilenkin1997}, including cosmic-string scenarios producing extremely high-energy neutrinos \cite{BerezinskyCosmicStrings2011}.

Experimentally, this energy domain is unusual because no single detection technique covers it completely.  The idea of detecting high-energy neutrinos with Cherenkov detectors deep in a lake or sea goes back to Markov \cite{Markov1960}, while coherent radio emission from high-energy cascades in dense media was identified by Askaryan \cite{Askaryan1962}.  Optical-Cherenkov neutrino telescopes provide the largest exposures at the lower end of the range, air-shower experiments become competitive around EeV energies, and radio techniques extend the reach by many further decades in energy.  The observation of a 220-PeV event KM3-230213A by KM3NeT \cite{KM3NeT:Nature} have renewed interest in this energy region.  At the same time, most searches above the PeV scale remain statistically limited, and the available information is distributed among many experiments using very different detector concepts.

This diversity makes a direct comparison of published upper limits less straightforward than a collection of curves on a common plot may suggest.  Different analyses quote single-flavor or all-flavor fluxes, sometimes separately or jointly for neutrinos and antineutrinos; they use different logarithmic energy intervals and test spectra; and published sensitivities and observed limits rely on different confidence constructions, event counts, and background treatments.  A consistent combination therefore requires returning, as far as the public information permits, to effective areas, exposures, and counting inputs rather than combining the plotted flux limits themselves.

UHE-neutrino flux scenarios and their observational constraints have been studied for decades, see e.g.\ Refs.~\cite{SemikozUHE1,SemikozUHE2,Ryabov:2006pk,Kotera:2025jca}.  The broader experimental landscape is reviewed in the Snowmass white paper \cite{SnowmassUHE}, while discovery prospects for next-generation detectors have been quantified, for example, in Ref.~\cite{ValeraUHE2023}.  Recent work has also stressed the assumptions involved in converting an incident neutrino flux into expected event yields through detector effective areas \cite{Palmisano2026}, while multi-experiment studies motivated by KM3-230213A illustrate the growing interest in joint interpretations of the available UHE-neutrino data \cite{ClashTitans2026,KM3NeT:agnostic,BaikalGVD2025}.  The purpose of the present study is complementary: we construct an exposure-level combination of published diffuse searches in a single statistical and flavor convention.

We focus on the isotropic diffuse neutrino flux at energies between $10^{16}$~eV and $10^{26}$~eV.  We retain published analyses for which an energy-dependent exposure, or an equivalent effective area and live time, can be reconstructed with sufficient accuracy and whose contribution is non-negligible relative to the other searches over the energy interval in which they overlap.  All effective areas and exposures are converted to a common total all-flavor $\nu+\bar\nu$ convention assuming $\nu_e:\nu_\mu:\nu_\tau=1:1:1$ at Earth.  The reconstructed exposures are combined with the published numbers of surviving events and expected backgrounds in a uniform Poisson construction.  Our primary result is a set of 90\% C.L. quasi-differential upper bounds for one-decade $E_\nu^{-1}$ test spectra.  The same exposures are then used to constrain fixed theoretical spectra by direct folding rather than by comparison with the quasi-differential points.

The construction is deliberately limited to information that can be treated uniformly across experiments.  It is not intended to replace the dedicated likelihood analyses of the individual collaborations, which may exploit reconstructed energies, directions, event classifications, and detector-specific systematics unavailable in public form.  Its purpose is instead to provide a transparent common baseline that can be reproduced from published information and applied consistently to models spanning energy ranges much broader than the sensitivity of any one experiment.

The experimental inputs are summarized in Sec.~\ref{sec:data}.  Section~\ref{sec:limits} describes the common exposure convention and the statistical combination; it also presents the quasi-differential limit.  In Sec.~\ref{sec:disc}, we apply the result to representative cosmogenic-neutrino and exotic spectra and discuss projected sensitivities.  Details of the exposure reconstruction and of the projected-sensitivity conversions are collected in the appendices.

\section{Experimental inputs}
\label{sec:data}

The included searches span optical-Cherenkov neutrino telescopes, air-shower arrays, radio detectors of polar ice, and lunar radio observations.  For each analysis we retain the published number of surviving events and expected background whenever these quantities affect the count-level combination described in Sec.~\ref{sec:limits}.

\subsection{Neutrino telescopes}
\label{sec:data:NT}

\textit{IceCube.}  We use the recent IceCube search for extremely high-energy neutrinos based on 12.6 years of data \cite{IceCube2025}.  It extends the optical-telescope exposure well above the PeV range and provides the dominant optical contribution around EeV energies.

\textit{Baikal-GVD.}  For Baikal-GVD we use the dedicated multi-PeV diffuse-flux analysis based on cascade-like events, which quotes limits over $10^{15.5}$--$10^{20}$~eV \cite{BaikalGVD2025}.  Its contribution is concentrated near the lower-energy boundary of the interval considered here.

\textit{KM3NeT.}  KM3NeT has reported the ultra-high-energy event KM3-230213A \cite{KM3NeT:Nature}.  We use the exposure information associated with the model-independent placement of this event in the global UHE-neutrino landscape \cite{KM3NeT:agnostic}; the non-zero event count is retained in the statistical treatment of Sec.~\ref{sec:limits}.

Earlier optical diffuse searches were also checked when defining the data set.  The AMANDA-II UHE search  \cite{AMANDAIIUHE} and the Baikal NT200 \cite{BaikalNT200} diffuse search overlap the lower part of our energy range but are superseded there by the much larger IceCube and Baikal-GVD exposures; the nine-year ANTARES all-flavor diffuse analysis is concentrated mainly below the $10^{16}$~eV lower boundary chosen here \cite{ANTARESDiffuse}.  We therefore do not include these legacy optical data sets numerically.

\subsection{Cosmic-ray experiments}
\label{sec:data:CR}

\textit{Pierre Auger Observatory.}  The Pierre Auger Observatory is sensitive to neutrinos through very inclined, deeply developing air showers.  We include both the downward-going channel, sensitive to all flavors, and the Earth-skimming $\nu_\tau$ channel, whose use for UHE-neutrino searches was developed early in Ref.~\cite{Bertou2002}, using the published diffuse-neutrino exposure through August 2018 and its later update through the end of 2021 \cite{Auger:effarea,Auger:limit2022}.  Around EeV energies, Auger provides one of the largest non-radio contributions to the combined exposure.

The High Resolution Fly's Eye (\textit{HiRes}) \cite{HiRes-neutrino}, the \textit{Telescope Array} experiment \cite{TelescopeArray-neutrino}, and the balloon-borne EUSO-SPB2 Cherenkov Telescope \cite{EUSOSPB2} have also reported diffuse-neutrino constraints. While Refs.~\cite{HiRes-neutrino,TelescopeArray-neutrino} do not provide required numerical information about exposures, Ref.~\cite{EUSOSPB2} presents a short-duration proof-of-principle result.  In each of the three cases, the exposure is too small to make a visible contribution to the combined bound in the energy range where more sensitive searches already enter; we therefore do not include these results numerically.

\subsection{Radio detection: observing the Earth}
\label{sec:data:radio-Earth}

At energies $\gtrsim 10^{17}$~eV, coherent radio emission from particle cascades provides access to effective volumes much larger than those available to optical techniques.  The experiments below search primarily for Askaryan emission from neutrino-induced showers in polar ice; ANITA also provides a distinct channel for upward-going air showers initiated by tau leptons.

\textit{FORTE.}  The FORTE satellite searched for impulsive radio emission from cascades in the Greenland ice sheet.  Its neutrino analysis corresponds to an effective observing time of about three days accumulated between 1997 and 1999 \cite{FORTE}.  One uncertain event survived the final selection.  Although the paper describes it as background noise, the published flux limit uses the one-event Poisson upper count $s_{\rm up}=3.89$ without introducing a separate expected background mean.  We therefore retain $(n_i,b_i)=(1,0)$ in the combination.

\textit{RICE.}  The Radio Ice Cherenkov Experiment operated antennas embedded in South Pole ice.  We use its final diffuse-flux analysis, based on data taken between 1999 and 2010, together with the published energy-dependent acceptance \cite{RICE}.  No neutrino candidate from this search enters the combined count.

\textit{ANITA.}  We treat separately the in-ice Askaryan searches from ANITA I--IV \cite{ANITA1,ANITA2,ANITA3,ANITA-Askaryan} and the upward-going $\nu_\tau$ air-shower acceptance studied for ANITA-IV \cite{ANITA-tau}.  The four full-flight Askaryan searches contain three surviving candidates in total, $0+1+1+1$, with a summed expected background $0+0.97+0.70+0.64=2.31$ events; for ANITA-II we use the corrected event count given in the published erratum to Ref.~\cite{ANITA2}.  The earlier ANITA-lite exposure \cite{ANITAlite} is retained as a zero-count precursor data set.  The four ANITA-IV near-horizon air-shower events are different statistically.  Ref.~\cite{ANITA-tau} finds them observationally compatible with $\tau$-induced showers, while a diffuse isotropic Standard-Model interpretation is in strong tension with the Auger constraints and, at the highest energies, with the ANITA Askaryan channel. At the same time, Ref.~\cite{ANITA-tau} does not define an internal diffuse-search background expectation for these four events.  To avoid importing the external comparison back into the combined likelihood, the upward-going channel is  omitted from the primary statistical combination.  We retain both its reconstructed exposure and all four events in two robustness variants: an all-candidate treatment, $(n_i,b_i)=(4,0)$ for this channel, and the phenomenological background-tagged treatment, $(n_i,b_i)=(4,4)$. The corresponding exposure and flavor conversion are specified in \ref{sec:app:details}.

\textit{ARIANNA.}  We use the diffuse search with the seven-station ARIANNA pilot array, based on data collected from December 2014 to February 2019 and corresponding to 2906.9 station-days of livetime \cite{ARIANNA}.  The collaboration used zero background to construct the published quasi-differential sensitivity; we retain that convention only when inverting the sensitivity to recover the exposure.  In the combined count we use zero surviving events and the corresponding one-decade background estimate $b_i\simeq0.08$.

\textit{ARA.}  For the observational combination we use the published four-year analysis of ARA stations A2 and A3 \cite{ARA}, with zero surviving events and an expected residual background of $(5\pm2)\times10^{-2}$ events per station.  Treating A2 and A3 as one combined ARA channel therefore gives $b_i=0.10$ in the likelihood.  The published flux limit itself used the zero-background Feldman--Cousins count, which we retain only when reconstructing the exposure from that limit.  A much larger 2013--2023 full-array trigger-level sensitivity has subsequently been presented \cite{ARAAsBuilt}, and the first array-wide diffuse search is in progress \cite{ARAArrayWideStatus2026}.  Because no corresponding final diffuse-flux upper limit has yet been published, these newer results are discussed only with the projections in Sec.~\ref{sec:prospects}.

\subsection{Radio detection: observing the Moon}
\label{sec:data:radio-Moon}

Lunar-Askaryan searches use the lunar regolith as a neutrino target and search from Earth for coherent radio pulses produced by interactions near the lunar surface \cite{DagkesamanskiiZheleznykh1989}.  Their thresholds are high, but the enormous target volume extends the exposure to the highest energies considered here.  Several early searches published flux limits rather than tabulated exposures; for these data sets we combine the published observing times with later calculations or compilations of the corresponding effective apertures \cite{JamesProtheroe,LUNASKA-Parkes}.

\textit{GLUE.}  The Goldstone Lunar Ultra-high Energy neutrino Experiment used antennas of the NASA/JPL Deep Space Network and accumulated approximately 120~h of lunar observations \cite{GLUE,JamesProtheroe}.

\textit{Parkes and Kalyazin.}  Early lunar-Cherenkov searches were performed with the 64-m Parkes telescope \cite{Parkes1996} and the 64-m antenna of the Kalyazin Radio Astronomical Observatory \cite{Kalyazin}.  Their exposures are reconstructed using the effective-aperture calculation of Ref.~\cite{JamesProtheroe}.

\textit{RESUN.}  Project RESUN used the Expanded Very Large Array for about 200~h of lunar observations \cite{RESUN}.  We use the RESUN effective-aperture curve summarized in Fig.~3 of Ref.~\cite{LUNASKA-Parkes}, retaining the experiment-specific aperture calculation underlying that curve.

\textit{LUNASKA--ATCA.}  The LUNASKA search with the Australia Telescope Compact Array accumulated 26.15~h of on-Moon observations \cite{LUNASKA-ATCA}.  We use the LUNASKA--ATCA effective-aperture curve summarized in Fig.~3 of Ref.~\cite{LUNASKA-Parkes}.

\textit{NuMoon--WSRT.}  The NuMoon experiment used the Westerbork Synthesis Radio Telescope to search for nanosecond radio pulses from the Moon.  The data set entering the published limit contains 46.7~h of observation \cite{WSRT-2009,WSRT2010}.

\textit{LUNASKA--Parkes.}  The later LUNASKA observations with Parkes provide both the flux limit and the corresponding effective aperture directly \cite{LUNASKA-Parkes}.  This is the most recent of the historical lunar-Cherenkov data sets included in the combination.

\begin{figure}
\centering
\includegraphics[width=0.75\linewidth]{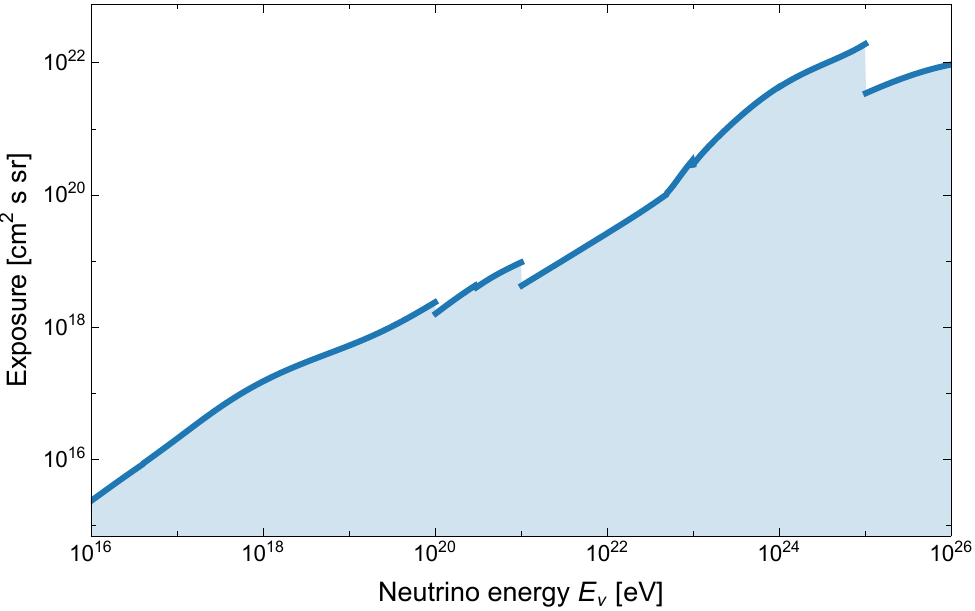}\\[3ex]
\includegraphics[width=0.75\linewidth]{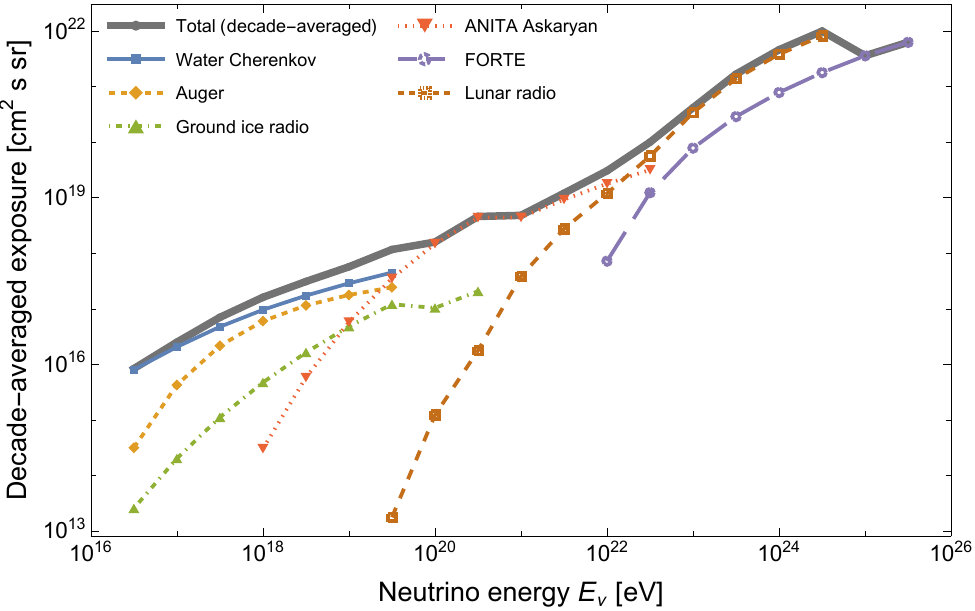}
\caption{Combined exposure to the total all-flavor diffuse cosmic-neutrino flux, assuming a $1{:}1{:}1$ flavor ratio at Earth.  \textbf{Top:} pointwise sum of the published exposures within their reported energy ranges.   \textbf{Bottom:} logarithmic one-decade average exposures of the main detector classes.  A channel contributes only when the upper edge of the one-decade test-spectrum window lies within its published/reconstructed exposure range. The markers correspond to evaluated bin-by-bin values; connecting lines only guide the eye.  The thick gray curve connects bin-by-bin total exposure values in the same convention.}\label{fig:plot1}
\end{figure}

The exposures of all included searches are converted to the common all-flavor convention and combined in Sec.~\ref{sec:limits}.  Figure~\ref{fig:plot1} shows both their pointwise sum and one-decade averages grouped by detection method.  The decomposition makes the complementarity explicit: water and ice Cherenkov telescopes dominate the low-energy end, Auger and radio searches become important around the EeV scale, ANITA controls part of the intermediate radio range, and lunar observations extend the exposure coverage to the highest energies.

\section{Limits on the neutrino flux}
\label{sec:limits}

The data sets are combined through their energy-dependent exposure to a common diffuse flux.  The exposure $\mathcal{E}_i(E_\nu)$ of each amalysis $i$ determines the expected signal count,
\begin{equation}
 s_i=\int dE_\nu\,\mathcal{E}_i(E_\nu)\,
 \frac{d\Phi}{dE_\nu}.
 \label{eq:mu-exposure}
\end{equation}
Here $E_\nu$ is the neutrino energy and $d\Phi/dE_\nu$ is the differential flux per unit solid angle in the convention specified below.  All exposures are expressed for the total all-flavor $\nu+\bar\nu$ flux, assuming the flavor ratio $\nu_e:\nu_\mu:\nu_\tau=1:1:1$ at Earth.  In particular, when a published effective area is summed over the three flavors but is folded in the source analysis with an equal per-flavor flux, the effective area used with the total all-flavor flux is one third of that sum.  We denote by $\mathcal{P}$ the primary channel set, excluding the ANITA-IV upward-going channel used only in robustness tests.  For visualization and fixed-spectrum folding we also define the sum of its available exposure pieces,
\begin{equation}
 \mathcal{E}(E_\nu)=\sum_{i\in\mathcal{P}} \mathcal{E}_i(E_\nu),
 \label{eq:total-exposure}
\end{equation}
where each term is used only over the energy interval for which that exposure is published or reconstructed. This function $\mathcal{E}_(E_\nu)$ is presented in the upper panel of Rig.~\ref{fig:plot1} A numerical table of $\mathcal{E}_(E_\nu)$ on the 0.01-decade grid is provided with this publication as the ancillary file \texttt{total\_exposure.txt}.  Before integration, tabulated or digitized effective-area and exposure curves are replaced by smooth functions of log-energy; the fit form, coefficients, and validation are given in \ref{sec:app:smoothfits}.

As it is customary in the field, we quote quasi-differential bounds in one-decade-wide intervals.  For a bin centered at $E_0$,
\begin{equation}
 E_0/\sqrt{10}<E_\nu<\sqrt{10}\,E_0,
\end{equation}
we use an $E_\nu^{-1}$ test spectrum,
\begin{equation}
 \frac{d\Phi}{dE_\nu}=\frac{Q(E_0)}{E_0E_\nu},\qquad
 Q(E_0)\equiv E_0^2\left.\frac{d\Phi}{dE_\nu}\right|_{E_0},
 \label{eq:bin-spectrum}
\end{equation}
and set it to zero outside the interval.  
The dimension of $Q$ is $\mathrm{GeV\,cm^{-2}\,s^{-1}\,sr^{-1}}$.  
For this test spectrum, the expected signal contribution of analysis $i$ is
\begin{equation}
 s_i(Q)=Q\,\alpha_i(E_0),\qquad
 \alpha_i(E_0)=\frac{1}{E_0}
 \int_{E_0/\sqrt{10}}^{\sqrt{10}E_0}
 \mathcal{E}_i(E_\nu)\,\frac{dE_\nu}{E_\nu}.
 \label{eq:alpha-bin}
\end{equation}
The energy-dependent exposure is thus integrated over the full decade; no bin-center approximation is used.  A particular analysis is included in a bin only if its published exposure covers the entire bin. For the combined ANITA Askaryan channel this support test is applied flight by flight before the flight exposures are summed.

For each bin, let $\mathcal{P}(E_0)\subseteq\mathcal{P}$ be the subset of primary channels whose exposures cover the full one-decade interval.  We retain the observed count $n_i$ and expected background $b_i$ of every $i\in\mathcal{P}(E_0)$ and use the channel-wise likelihood below, where ${\rm Pois}(n\mid\mu)$ denotes the Poisson probability mass function with mean $\mu$:
\begin{equation}
 \mathcal{L}(Q)=\prod_{i\in\mathcal{P}(E_0)}
 {\rm Pois}\!\left(n_i\,\middle|\,Q\alpha_i(E_0)+b_i\right).
 \label{eq:product-poisson}
\end{equation}
The best-fit amplitude $\widehat Q\ge0$ maximizes Eq.~\eqref{eq:product-poisson}.  For a tested amplitude $Q$ we use the one-sided profile-likelihood statistic
\begin{equation}
 q_Q=
 \begin{cases}
 -2\ln\!\left[\mathcal{L}(Q)/\mathcal{L}(\widehat Q)\right], & \widehat Q\le Q,\\
 0, & \widehat Q>Q.
 \end{cases}
 \label{eq:qQ}
\end{equation}
The published $b_i$ are treated as fixed expectations.  
All active channels with $b_i=0$ can be compressed exactly into a single sufficient channel.  
With all sums restricted to $\mathcal{P}(E_0)$, define $R_0=\sum_{b_i=0}\alpha_i(E_0)/\sum_j\alpha_j(E_0)$, $N_0=\sum_{b_i=0}n_i$, and $t=Q\sum_j\alpha_j(E_0)$.  
Their dependence on the tested signal normalization is proportional to $\exp(-tR_0)t^{N_0}$; the allocation of the $N_0$ events among the zero-background channels contributes only a multiplicative factor independent of $Q$.
After this compression, the primary likelihood has at most four Poisson factors.  Let $q_Q^{\rm obs}$ denote the statistic evaluated on the observed counts and $q_Q^{\rm toy}$ its value for a Poisson realization generated at the tested $Q$.  For $p_Q=P(q_Q^{\rm toy}\ge q_Q^{\rm obs}\mid Q)$, the quoted upper limit is defined as
\begin{equation}
 Q_{90}(E_0)=\inf\{Q:p_Q\le0.10\} .
 \label{eq:combined-limit}
\end{equation}
This construction retains which experiment contains each surviving event and therefore uses more of the public counting information than an aggregate Poisson count.  It still remains a count-level likelihood: reconstructed event energies, directions, classifier variables, and collabo\-ration-specific nuisance parameters are not available uniformly and are not used.

The non-zero count and background inputs are summarized in Table~\ref{tab:event-backgrounds}.  The ANITA-IV upward-going channel is omitted from the primary likelihood for the reason given in Sec.~\ref{sec:data:radio-Earth}; we nevertheless repeat the calculation with $(n_i,b_i)=(4,0)$ for this channel and with the phenomenological assignment $(n_i,b_i)=(4,4)$.  The latter changes the primary one-decade bounds by at most about 11\%, whereas treating all four events as possible diffuse candidates weakens the limit by as much as a factor $2.5$ where the upward-going exposure is active.

\begin{figure}[!htb]
\centering
\includegraphics[width=0.72\linewidth]{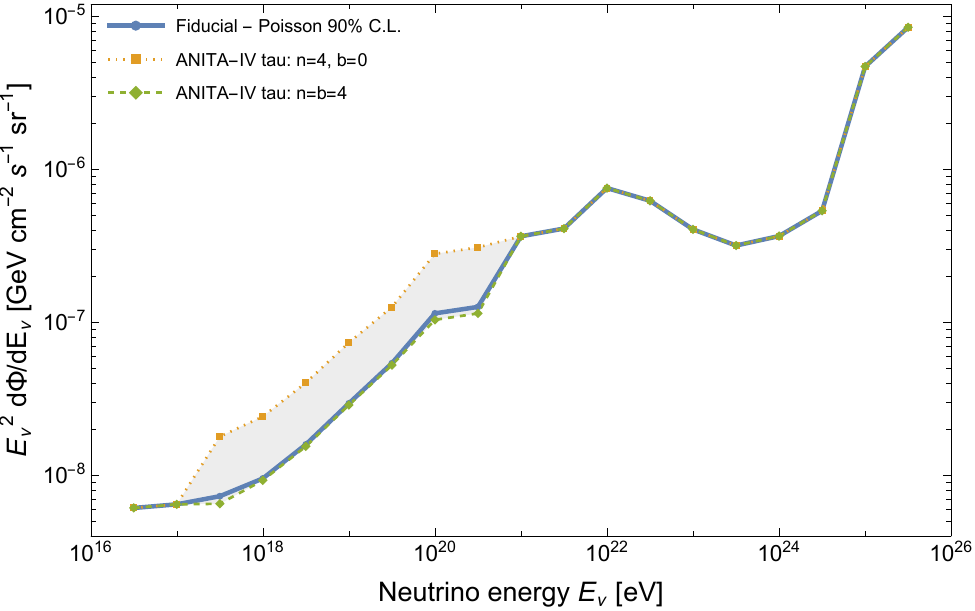}
\caption{Combined 90\% C.L. one-decade upper bounds on the isotropic total all-flavor diffuse neutrino flux.  The solid curve is the fiducial channel-wise product-Poisson bound. The shaded envelope brackets the treatment of four ANITA-IV near-horizon events, with $n_i=4$ and $0 \le b_i \le 4$ for that channel.}
\label{fig:combined-limit}
\end{figure}

We evaluate the bound at half-decade spacing, $E_0=10^{16.5},10^{17},\ldots,10^{25.5}$~eV.  Adjacent points therefore correspond to overlapping one-decade intervals and are strongly correlated.  The numerical values are listed in Table~\ref{tab:combined-limits} and presented in Fig.~\ref{fig:combined-limit}.  The strongest primary bound is
$E_\nu^2d\Phi/dE_\nu=6.1\times10^{-9}\,
\mathrm{GeV\,cm^{-2}\,s^{-1}\,sr^{-1}}$ in the lowest energy bin. 

\section{Discussion}
\label{sec:disc}

The energy-dependent exposure can be used to constrain the normalization of any fixed spectral shape.  We denote the differential model flux by $\phi_{\rm model}(E_\nu)\equiv d\Phi_{\rm model}/dE_\nu$ and introduce a dimensionless scale factor $k$ through
\begin{equation}
\frac{d\Phi}{dE_\nu}=k\,\phi_{\rm model}(E_\nu).
\end{equation}
For each channel we compute the nominal ($k=1$) signal
\begin{equation}
s_i(k=1)=\int dE_\nu\,\mathcal{E}_i(E_\nu)\,\phi_{\rm model}(E_\nu).
\label{eq:model-folding}
\end{equation}
Let $\mathcal{P}_{\rm model}\subseteq\mathcal{P}$ contain the primary channels whose published exposure overlaps the tabulated model range, and define the total nominal yield $S(k=1)\equiv\sum_{i\in\mathcal{P}_{\rm model}}s_i(k=1)$.  The channel-wise likelihood is then
\begin{equation}
\mathcal{L}(k)=\prod_{i\in\mathcal{P}_{\rm model}} {\rm Pois}\!\left(n_i\,\middle|\,k s_i(k=1)+b_i\right),
\label{eq:model-product}
\end{equation}
and the 90\% endpoint $k_{90}$ is obtained with the same one-sided likelihood-ratio statistic as in Eq.~\eqref{eq:qQ}, calibrated for the fixed spectrum with deterministic Poisson Monte Carlo.  We impose $E_\nu\ge10^{16}$~eV and do not extrapolate a theory curve beyond the energy support supplied by its source.  Again, each detector contributes only when  its published exposure overlaps with the theoretical model energy range; any unreported high-energy continuation of the detector exposure is omitted. This makes the expected signal, and hence the resulting normalization constraint, conservative with respect to the unknown continuation of the exposure.  A value $k_{90}<1$ means that the nominal fixed normalization lies above the 90\% endpoint of this count-level test.  Note that when the published benchmark model was itself fitted using some of the same neutrino data, this is a compatibility test of that fixed benchmark with the broader compilation and not an independent exclusion of a pre-data model.

\subsection{Cosmogenic neutrino models}
\label{sec:disc:cosmogenic}
\begin{figure}[!htb]
\centering
\includegraphics[width=0.78\linewidth]{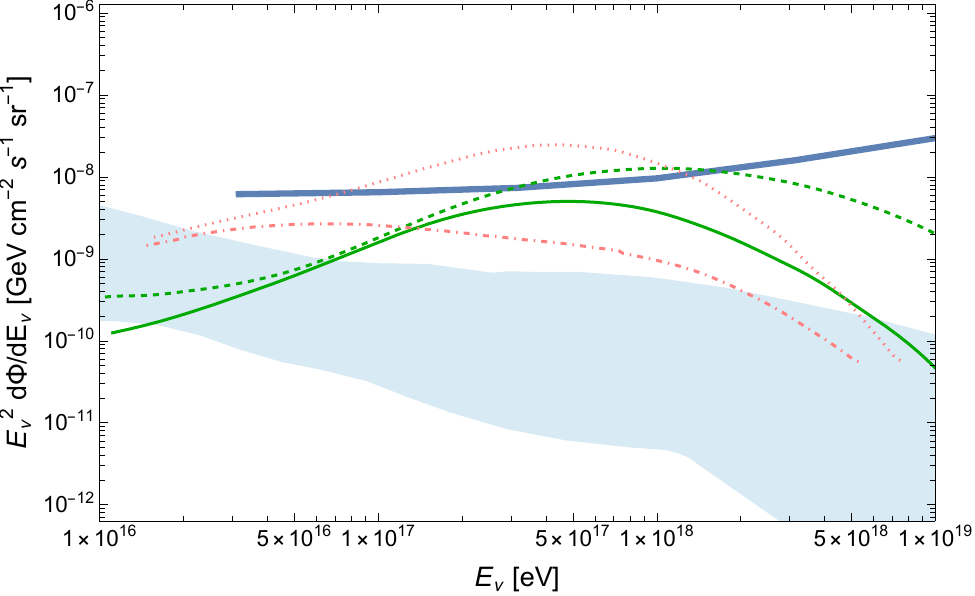}
\caption{The combined 90\% C.L. one-decade flux limit (thick blue line)  and representative cosmogenic-neutrino predictions: the Auger composition-constrained band \cite{AugerUHECR2023} (shaded), Telescope-Array best-fit (pink dotted) and local-minimum (pink dash-dotted) \cite{KuznetsovCosmogenic2026}, and KM3NeT-only (green dashed) and joint (green full) fits of Ref.~\cite{AlhebsiCosmogenic2026}. All curves are shown as total all-flavor $E_\nu^2\,d\Phi_\nu/dE_\nu$.}
\label{fig:cosmogen}
\end{figure}

Figure~\ref{fig:cosmogen} compares the combined limit with six representative cosmogenic-neutrino predictions.  The shaded band presents the range of models associated with the Pierre Auger fits of the UHECR spectrum and composition \cite{AugerUHECR2023}.  The Telescope Array benchmarks are the best fit and the local minimum of the minimal UHECR source model from Ref.~\cite{KuznetsovCosmogenic2026}.  Finally, Ref.~\cite{AlhebsiCosmogenic2026} studies a two-population UHECR model containing a mixed-composition population and a subdominant UHE-proton population.  Its Fig.~2 gives a KM3NeT-only fit and a joint fit including the null IceCube and Pierre Auger observations; within each fit we sum both the mixed-composition and UHE-proton neutrino components before folding the total spectrum with the exposure.

The original curves are given per flavor and are multiplied by three to match our all-flavor convention.  The resulting 90\% normalization limits are collected in Table~\ref{tab:cosmogenic-k90}. The numbers for models of Ref.~\cite{AlhebsiCosmogenic2026} should be read as compatibility diagnostics because the benchmark spectra were inferred using overlapping neutrino data.  The ANITA-IV treatment is an issue for broad spectra: if all four near-horizon events are retained as possible diffuse candidates [$(n_i,b_i)=(4,0)$ for that channel], the corresponding limits become approximately $1.6$ for the TA best fit, $0.83$ for the KM3NeT-only fit, and $4.4$ for the joint fit.  The historical background-tagged $(n_i,b_i)=(4,4)$ variant, by contrast, differs from the primary values only at the few-percent level. 

\begin{table}[!htb]
\centering
\caption{Upper limits on the normalization of cosmogenic-neutrino benchmark spectra shown in Fig.~\ref{fig:cosmogen}.  $S$ is the total, energy-integrated expected signal yield over the primary channels that overlap the model range.  $k_{90}$ is the fiducial upper limit; the final column gives the conservative variant in which all four ANITA-IV near-horizon events are possible diffuse candidates.}
\label{tab:cosmogenic-k90}
\begin{tabular}{lccc}
\hline
Model & $S(k=1)$ & $k_{90}$ & $k_{90}^{\tau\,{\rm cand}}$ \\
\hline
Auger envelope, lower \cite{AugerUHECR2023} & 0.0529 & 66.1 & 168 \\
Auger envelope, upper \cite{AugerUHECR2023}& 1.406 & 2.54 & 6.35 \\
TA best fit \cite{KuznetsovCosmogenic2026}& 5.546 & 0.641 & 1.61 \\
TA local minimum \cite{KuznetsovCosmogenic2026} & 2.325 & 1.49 & 3.82 \\
KM3NeT-only total fit \cite{AlhebsiCosmogenic2026} & 10.8 & 0.325 & 0.830 \\
Joint total fit \cite{AlhebsiCosmogenic2026} & 2.023 & 1.74 & 4.42 \\
\hline
\end{tabular}
\end{table}

\subsection{Heavy dark matter and topological defects}

If neutrinos are produced in decays or annihilation of superheavy dark-matter particles, the nonuniform distribution of dark matter in the Milky-Way halo results in a significant anisotropy of the expected signal over the sky. These models cannot be tested consistently with the direction-averaged diffuse fluxes we study here.  Writing the directional differential intensity as $\phi_{\rm DM}(E_\nu,\Omega)\equiv d^2\Phi_{\rm DM}/(dE_\nu\,d\Omega)$, with $\Omega$ denoting arrival direction, a consistent multi-experiment test would require
\begin{equation}
 s_i^{\rm DM}=\int dE_\nu\,d\Omega\,\mathcal{E}_i(E_\nu,\Omega)\,\phi_{\rm DM}(E_\nu,\Omega),
 \label{eq:dm-directional}
\end{equation}
where $\mathcal{E}_i(E_\nu,\Omega)$ is the direction-dependent exposure.  Such directional acceptances or exposures are not publicly available for every experiment in a uniform form.  We therefore do not quote a numerical normalization or lifetime constraint on the SHDM models and do not overlay them on the isotropic-limit figure; doing so would invite an interpretation that the present data reduction cannot support.

\begin{figure}[!htb]
\centering
\includegraphics[width=0.78\linewidth]{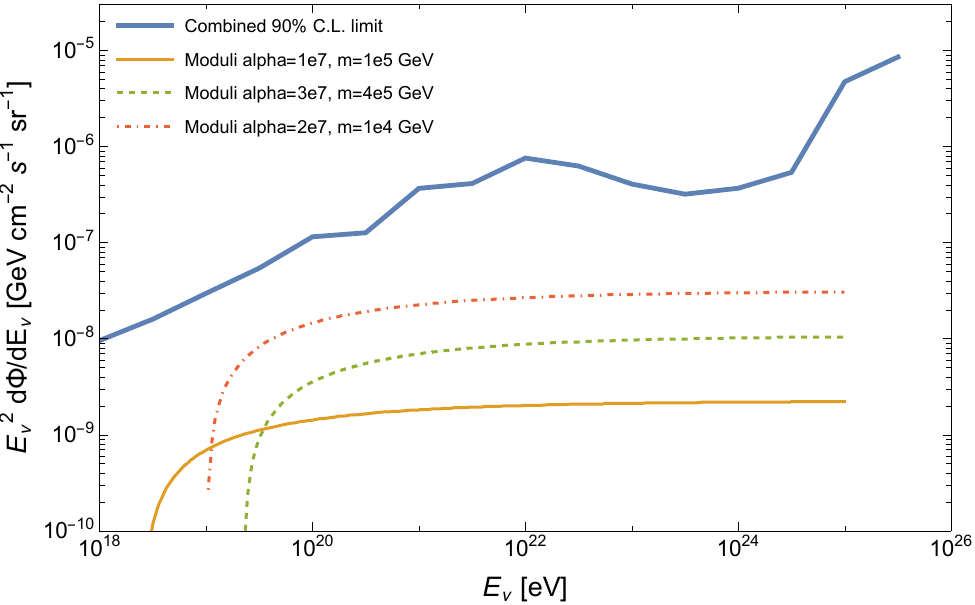}
\caption{The fiducial combined 90\% C.L. one-decade flux upper limit compared with three cosmic-string/modulus benchmarks considered in Ref.~\cite{PUEOSensitivity}, based on Ref.~\cite{BerezinskyCosmicStrings2011}.}
\label{fig:exotics}
\end{figure}

This is not the case for neutrinos produced in rare decays of cosmic topological defects, in which case the signal is dominated by distant objects. Figure~\ref{fig:exotics} shows three isotropic cosmic-string benchmarks used in Ref.~\cite{PUEOSensitivity}, based on modulus emission from cosmic-string cusps \cite{BerezinskyCosmicStrings2011}.  Denote the modulus--string coupling by $\alpha_{\rm cs}$ and the modulus mass by $m$.  For $(\alpha_{\rm cs},m)=(10^7,10^5\,{\rm GeV})$, $(3\times10^7,4\times10^5\,{\rm GeV})$, and $(2\times10^7,10^4\,{\rm GeV})$, the primary normalization endpoints are $k_{90}\simeq25$, $12$, and $3.4$, respectively; all three nominal benchmark spectra therefore remain well below the present constraint.  With the conservative ANITA-IV all-candidate treatment these become approximately $47$, $23$, and $6.2$.  Here, $k$ is only a multiplicative normalization of the fixed benchmark spectrum and cannot be converted into a simple bound on $\alpha_{\rm cs}$, because $\alpha_{\rm cs}$ also changes the energy-dependent minimum redshift and hence the spectral shape \cite{PUEOSensitivity}.

\subsection{Prospects}
\label{sec:prospects}

\begin{figure}[!htb]
\centering
\includegraphics[width=0.82\linewidth]{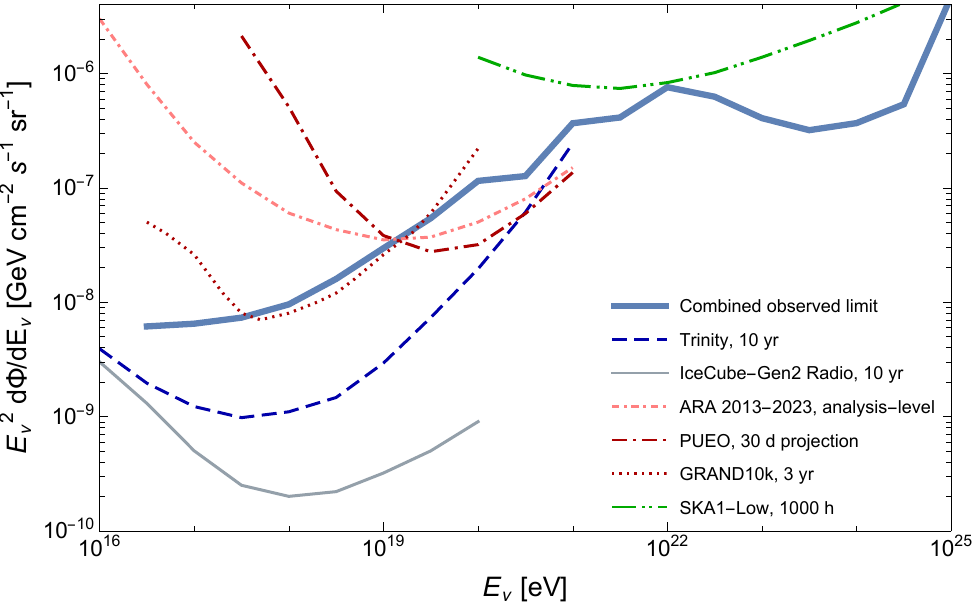}
\caption{Selected prospective sensitivities, shown for context and not included in the combined limit.  ARA and PUEO have already collected the relevant data, but their diffuse-flux limits are not yet published.  The GRAND10k curve corresponds to the optimistic (aggressive-threshold) three-year projection.  An extended compilation is shown in \ref{sec:app:prospects}.}
\label{fig:prospects-main}
\end{figure}

The combined limit presented above is constructed  from published observational results for which the information required for a uniform statistical treatment is available.  To indicate the likely near- and medium-term improvement, Fig.~\ref{fig:prospects-main} shows a deliberately sparse selection of projected sensitivities that are \emph{not} included in the combination (for a more representative list of proposed experiments, see \ref{sec:app:prospects}).  The selection spans the full energy interval and several complementary detection concepts; it is not intended as an exhaustive catalogue.  We show the full Trinity Observatory \cite{TrinitySensitivity}, the full-array ARA simplified analysis-level sensitivity \cite{ARAAsBuilt}, the published 30-day PUEO projection \cite{PUEOSensitivity,PUEOSimulation,PUEOWhitePaper}, the IceCube-Gen2 in-ice radio array \cite{IceCubeGen2Radio}, the GRAND10k stage \cite{GRANDScience}, and the lunar-Askaryan sensitivity of SKA1-Low \cite{SKALunar}.  Current project-status reports show that several of these concepts have meanwhile progressed beyond their original design studies, including the Trinity Demonstrator, the GRAND prototypes, and the first RNO-G neutrino search \cite{TrinityStatus2025,GRANDStatus2025,RNOGStatus2025}; the sensitivity curves themselves are nevertheless kept at their published reference configurations.

ARA and PUEO require a special comment.  In both cases the relevant data have already been taken, but a final diffuse-flux upper limit based on these data has not yet been published.  The ARA curve uses the 2013--2023 array exposure and a simplified event selection \cite{ARAAsBuilt,ARAArrayWideStatus2026}, while the PUEO curve is the published 30-day projection; the first PUEO flight took place in December 2025--January 2026 \cite{PUEOFlight}. 

All prospective curves are displayed in the same all-flavor $E_\nu^2 \, d\Phi/dE_\nu$ units whenever the published information permits such a conversion.  The conversion rules, caveats, and an extended set of projects are given in \ref{sec:app:prospects}.

\section{Conclusion}
\label{sec:concl}

We have combined published searches for diffuse neutrinos over $10^{16}$--$10^{26}$~eV by reconstructing their energy-dependent exposures and placing them in a common total all-flavor $\nu+\bar\nu$ convention.  The event counts and background expectations are retained channel by channel in a product-Poisson likelihood rather than collapsed into a single aggregate count.  A one-sided profile-likelihood statistic is calibrated directly from the Poisson sampling law after an exact sufficient-statistic compression of zero-background channels.  The equations and numerical tables in the paper, together with the two ancillary fixed-width tables of the combined exposure and primary quasi-differential limits, provide the numerical information distributed with this work.
Cherenkov telescopes, air-shower detectors, Antarctic radio experiments, and lunar searches provide complementary pieces of the constraint.

Direct folding of model cosmogenic spectra illustrates both the gain from the channel-wise likelihood and the limits of interpretation.  The fiducial upper limits on the model-flux normalizations are $k_{90}\simeq0.33$ for the KM3NeT-only benchmark of Ref.~\cite{AlhebsiCosmogenic2026}, $1.7$ for its joint-fit benchmark, and $0.64$ for the Telescope Array best-fit benchmark of Ref.~\cite{KuznetsovCosmogenic2026}.  The cosmic-string benchmarks considered here remain well below the present sensitivity.  No numerical SHDM constraint is obtained: the Galactic contribution is anisotropic and a meaningful test requires direction-dependent detector acceptances or exposures.

The present construction remains less informative than a collaboration-level global likelihood: it does not use reconstructed event energies or directions and treats published background expectations as fixed.  Its advantage is complementary.  The assumptions are explicit, the same convention is maintained over ten decades in energy, and arbitrary isotropic spectral models can be folded with the reconstructed channel exposures without relying on visual comparison with heterogeneous published curves.  The projected sensitivities collected in Sec.~\ref{sec:prospects} and \ref{sec:app:prospects} indicate substantial room for improvement in future, while the exposure-level framework provides a direct route for incorporating future measurements.

\section*{Acknowledgements}
The author is indebted to A.R.~Alhebsi for encouraging discussions. 
Publicly available large language models assisted in code debugging and language editing. The author reviewed their suggestions and takes full responsibility for the content of the article.

\section*{Conflicts of Interest}
The author declares no conflict of interest.

\appendix

\section{Exposure reconstruction and statistical details}
\label{sec:app:details}

\subsection{Common flux and exposure convention}

Throughout the reconstruction, every detector exposure is expressed with respect to the total all-flavor $\nu+\bar\nu$ flux.  For a flavor ratio $1:1:1$ at Earth, with $\ell=e,\mu,\tau$ labeling flavor,
\begin{equation}
 \frac{d\Phi_\ell}{dE_\nu}=\frac{1}{3}\frac{d\Phi_{\rm all}}{dE_\nu},
\end{equation}
so that for an effective area or exposure supplied as a sum over flavors the corresponding all-flavor exposure obeys
\begin{equation}
 \mathcal{E}_{\rm all}(E_\nu)=\frac{1}{3}
 \sum_{\ell=e,\mu,\tau}\mathcal{E}_\ell(E_\nu)
 \label{eq:allflavor-exposure}
\end{equation}
when the source convention folds that sum with an equal per-flavor flux.  This factor is required for the IceCube, Baikal-GVD, KM3NeT, and Auger.  An exposure that is explicitly for $\nu_\tau+\bar\nu_\tau$ only, such as the ANITA-IV upward-going exposure, is divided by three when it is expressed against the total all-flavor flux.  

All tabulated effective areas and exposures are represented only within the energy interval supported by the corresponding source.  Interpolation is carried out in $\log_{10}E_\nu$--$\log_{10}\mathcal{E}$ space and no effective area or exposure is extrapolated beyond a published endpoint.  For quasi-differential limits, this does not mean setting the unreported continuation to zero: a channel is excluded whenever the edge of the one-decade window exceeds its exposure endpoint.  For broad fixed-spectrum folds only the published overlap is integrated, yielding a conservative signal estimate.

The principal convention conversions entering the exposure reconstruction are summarized in Table~\ref{tab:exposure-conventions}.

\begin{table}[!htb]
\centering
\small
\caption{Principal convention conversions used in the exposure reconstruction.}
\label{tab:exposure-conventions}
\begin{tabular}{p{0.28\linewidth}p{0.62\linewidth}}
\hline
\noalign{\vskip 4pt}Input & Treatment \\
\noalign{\vskip 4pt}\hline
\noalign{\vskip 4pt}IceCube, Baikal-GVD, KM3NeT, Auger & Flavor-summed effective area or exposure divided by 3 when expressed against the total all-flavor flux for a $1:1:1$ flavor ratio. \\
\noalign{\vskip 4pt}
ANITA-IV Askaryan & Livetime 24.52~days. \\
\noalign{\vskip 4pt}ANITA-IV upward-going $\nu_\tau$ & Single-flavor tau exposure divided by 3; robustness variants only. \\
\noalign{\vskip 4pt}ARA & Published ordinate interpreted in its native $E_\nu\,d\Phi/dE_\nu$ convention; no additional power of $E_\nu$ is introduced. \\
\noalign{\vskip 4pt}ARIANNA & Published ordinate is $E_\nu^2\,d\Phi/dE_\nu$; the explicit factor of $E_\nu$ required in the exposure inversion is retained. \\
\noalign{\vskip 4pt}RICE and earlier ANITA curves & Native differential-limit inversion conventions, including the source bin-width factors, are retained. \\
\noalign{\vskip 4pt}Lunar searches & Effective aperture multiplied by on-Moon livetime; no additional normalization correction identified. \\
\hline
\end{tabular}
\end{table}

\subsection{Reconstruction of the individual exposures}

The inputs fall into three practical classes.  For IceCube, Baikal-GVD, KM3NeT, and the Pierre Auger Observatory an effective area or exposure is available directly.  The neutrino-telescope effective areas are multiplied by their livetimes and the angular normalization used by the source, followed by the flavor conversion in Eq.~\eqref{eq:allflavor-exposure}.  For Auger we use the published exposure through 31 August 2018 and scale it linearly with observing time to the end of 2021, matching the data period of the limit used here \cite{Auger:effarea,Auger:limit2022}.

Several radio experiments publish a quasi-differential flux bound rather than a machine-readable exposure.  In those cases we invert the statistical convention used by the source to recover the exposure before performing the present combination.  For ARA and ARIANNA the papers give an explicit relation between exposure and the one-decade limit \cite{ARA,ARIANNA}.  FORTE provides its neutrino sensitivity $\lambda(E_\nu)$ directly in tabular form; in the present notation this quantity is equivalent to an exposure.  For RICE and the ANITA Askaryan searches we use the corresponding native upper-count and bin-width convention to reconstruct the exposure.  The resulting exposure is then combined with the event count and background independently, as described below.  In particular, ARIANNA used zero background for its published quasi-differential curve, but the present statistical combination uses the estimated per-decade background after the exposure has been reconstructed.

The lunar-Askaryan searches are reconstructed from effective apertures and observing times.  For Parkes 1996, Kalyazin, and GLUE we use the recalculated apertures of Ref.~\cite{JamesProtheroe}.  For RESUN, LUNASKA--ATCA, and NuMoon we use the aperture curves collected in Fig.~3 of Ref.~\cite{LUNASKA-Parkes}, retaining the distinct experiment-specific aperture calculations represented there.  The two LUNASKA--Parkes observing configurations use the apertures published directly in Ref.~\cite{LUNASKA-Parkes} and are treated separately before being summed at the exposure level.

\subsection{Smooth representation of effective areas and exposures}
\label{sec:app:smoothfits}

The numerical integrations do not use the digitized or Monte-Carlo effective-area, aperture, or exposure points directly.  Instead, each curve is represented by a smooth function of energy before it is folded with a test spectrum.  This choice is intentional.  Even an exact digitization of a published curve retains small point-to-point fluctuations associated with finite Monte-Carlo statistics, while the underlying geometrical acceptance and exposure are expected to vary smoothly with energy except at genuine physical structures.  Fitting therefore prevents such fluctuations, or sub-pixel digitization noise, from propagating into the integrated limits.

All smooth fits used here can be written in the single form
\begin{equation}
 F(x)=a+b x+c x^2+d x^3+e x^4+f x^5
      +g\ln(x-x_0)+h[\ln(x-x_0)]^2 .
 \label{eq:fit-general}
\end{equation}
Coefficients not required for a particular data set are set to zero.  Thus the ordinary fits use $e=f=h=x_0=0$, while the IceCube fit has $g=h=0$ and non-zero $e$ and $f$.  The shifted-log lunar fits are represented by non-zero $x_0$, and the RESUN and NuMoon fits also use non-zero $h$.  We write $A_{\rm eff}$ for effective area, $(A\Omega)_{\rm eff}$ for effective aperture, and $\mathcal{E}$ for exposure; the fitted ordinate is logarithmic in all cases.  For IceCube, over the range actually entering this work, $10^{16}\le E_\nu/{\rm eV}\le10^{20}$, the fit is performed to $\log_{10}(A_{\rm eff}/{\rm m^2})$; its rms residual is $4.0\times10^{-3}$ in that logarithmic ordinate and the largest deviation of an individual tabulated point is $2.7\%$.  

Table~\ref{tab:direct-fit-coeff} gives the coefficients for inputs available directly as effective areas or exposures.  The convention is that $F$ equals the logarithm of the dimensionless quantity stated in the second column.  For Auger, $\mathcal{E}_{\rm src}$ denotes the published exposure through 31 August 2018, in ${\rm cm^2\,s\,sr}$, before the observation-time rescaling described above.  The last two columns give the neutrino-energy interval over which the fit is used in the present analysis.  The quoted precision is sufficient to reproduce the integrations at the accuracy relevant here.

\begin{table*}[!htb]
\centering
\footnotesize
\setlength{\tabcolsep}{1.0pt}
\renewcommand{\arraystretch}{1.08}
\caption{Smooth-fit coefficients for direct effective-area or exposure inputs in the common form of Eq.~\eqref{eq:fit-general}.  The fitted quantity is understood to enter $F$ through its base-10 logarithm in the units shown.  Within each coefficient column the upper and lower entries correspond to the two coefficients shown in the header; the final column gives the neutrino-energy interval over which the fit is used.}
\label{tab:direct-fit-coeff}
\begin{tabular}{@{}>{\raggedright\arraybackslash}p{0.095\textwidth}>{\centering\arraybackslash}p{0.115\textwidth}>{\centering\arraybackslash}p{0.10\textwidth}rrrrc>{\centering\arraybackslash}p{0.115\textwidth}@{}}
\toprule
Data set & Fitted quantity & $x$ & \shortstack{$a$\\$f$} & \shortstack{$b$\\$g$} & \shortstack{$c$\\$h$} & \shortstack{$d$\\$x_0$} & $e$ & $E_\nu$ range (eV) \\
\midrule
Baikal-GVD $\nu_e$ & $A_{\rm eff}$ ($\rm m^2$) & \shortstack{$\log_{10}$\\$(E_\nu/{\rm PeV})$} & \shortstack{2.24149\\0} & \shortstack{-0.388771\\0.879091} & \shortstack{0.135960\\0} & \shortstack{-0.0113384\\0} & 0 & $10^{16}$--$10^{20}$ \\
\addlinespace[2.5pt]
Baikal-GVD $\nu_\mu$ & $A_{\rm eff}$ ($\rm m^2$) & \shortstack{$\log_{10}$\\$(E_\nu/{\rm PeV})$} & \shortstack{1.62925\\0} & \shortstack{-0.242646\\1.32168} & \shortstack{0.0895877\\0} & \shortstack{-0.00753364\\0} & 0 & $10^{16}$--$10^{20}$ \\
\addlinespace[2.5pt]
Baikal-GVD $\nu_\tau$ & $A_{\rm eff}$ ($\rm m^2$) & \shortstack{$\log_{10}$\\$(E_\nu/{\rm PeV})$} & \shortstack{1.56844\\0} & \shortstack{0.0695596\\0.834208} & \shortstack{0.0527076\\0} & \shortstack{-0.00556204\\0} & 0 & $10^{16}$--$10^{20}$ \\
\addlinespace[2.5pt]
KM3NeT & $A_{\rm eff}$ ($\rm m^2$) & \shortstack{$\log_{10}$\\$(E_\nu/{\rm GeV})$} & \shortstack{-198.115\\0} & \shortstack{-189.384\\562.018} & \shortstack{10.6774\\0} & \shortstack{-0.266591\\0} & 0 & \shortstack{$10^{16}$--$10^{20}$} \\
\addlinespace[2.5pt]
IceCube & $A_{\rm eff}$ ($\rm m^2$) & \shortstack{$\log_{10}$\\$(E_\nu/{\rm eV})$\\$-18$} & \shortstack{3.70881\\-0.00172024} & \shortstack{0.584701\\0} & \shortstack{-0.111159\\0} & \shortstack{0.0158480\\0} & 0.00402733 & $10^{16}$--$10^{20}$ \\
\addlinespace[2.5pt]
Auger & $\mathcal{E}_{\rm src}$ ($\rm cm^2\,s\,sr$) & \shortstack{$\log_{10}$\\$(E_\nu/{\rm GeV})$} & \shortstack{-555.729\\0} & \shortstack{-309.336\\1088.51} & \shortstack{14.7301\\0} & \shortstack{-0.312664\\0} & 0 & \shortstack{$4\!\times\!10^{16}$--\\$3\!\times\!10^{20}$} \\
\addlinespace[2.5pt]
ANITA-IV Askaryan & $(A\Omega)_{\rm eff}$ ($\rm km^2\,sr$) & \shortstack{$\log_{10}$\\$(E_\nu/{\rm eV})$\\$-17$} & \shortstack{-3.79656\\0} & \shortstack{0.973646\\0.793082} & \shortstack{0.403035\\0} & \shortstack{-0.0846885\\0} & 0 & $10^{18}$--$10^{21}$ \\
\bottomrule
\end{tabular}
\end{table*}

For exposures reconstructed from published limits we use the same smoothing principle.  In Table~\ref{tab:reconstructed-fit-coeff}, $x=\log_{10}(E_\nu/{\rm eV})-q_E$ and the fitted quantity is $\log_{10}(\mathcal{E}/\mathcal{E}_0)$, with $\mathcal{E}_0$ in ${\rm cm^2\,s\,sr}$.  All rows have $e=f=h=x_0=0$; the explicit zero columns are retained to make the relation to Eq.~\eqref{eq:fit-general} unambiguous.

\begin{table*}[!htb]
\centering
\footnotesize
\setlength{\tabcolsep}{3.6pt}
\renewcommand{\arraystretch}{1.08}
\caption{Smooth-fit coefficients for radio exposures reconstructed from published limits, using Eq.~\eqref{eq:fit-general}.  Here $x=\log_{10}(E_\nu/{\rm eV})-q_E$ and the fitted quantity is $\log_{10}(\mathcal{E}/\mathcal{E}_0)$, with $\mathcal{E}_0$ in ${\rm cm^2\,s\,sr}$.  Within each coefficient column the upper and lower entries correspond to the two coefficients shown in the header.}
\label{tab:reconstructed-fit-coeff}
\begin{tabular}{@{}lccrrrrr>{\centering\arraybackslash}p{0.16\textwidth}@{}}
\toprule
Data set & $q_E$ & $\mathcal{E}_0$ & \shortstack{$a$\\$f$} & \shortstack{$b$\\$g$} & \shortstack{$c$\\$h$} & \shortstack{$d$\\$x_0$} & $e$ & $E_\nu$ range (eV) \\
\midrule
ANITA-lite & 18 & ${10^{10}}$ & \shortstack{3.57831\\0} & \shortstack{0.301086\\2.11505} & \shortstack{0.109063\\0} & \shortstack{-0.0185489\\0} & 0 & $3\!\times\!10^{18}$--$10^{23}$ \\
\addlinespace[2.5pt]
ANITA-I & 17 & ${10^{10}}$ & \shortstack{0.508365\\0} & \shortstack{2.24342\\3.49597} & \shortstack{-0.565151\\0} & \shortstack{0.0423938\\0} & 0 & $10^{18}$--$10^{23}$ \\
\addlinespace[2.5pt]
ANITA-II & 17 & ${10^{10}}$ & \shortstack{5.81011\\0} & \shortstack{-3.32735\\8.22184} & \shortstack{0.358121\\0} & \shortstack{-0.0183435\\0} & 0 & $10^{18}$--$10^{23}$ \\
\addlinespace[2.5pt]
ANITA-III & 17 & ${10^{10}}$ & \shortstack{0.182939\\0} & \shortstack{2.97346\\1.89564} & \shortstack{-0.524013\\0} & \shortstack{0.0285900\\0} & 0 & $10^{18}$--$10^{21}$ \\
\addlinespace[2.5pt]
ANITA-IV upward $\nu_\tau$ & 17 & ${10^{10}}$ & \shortstack{6.43671\\0} & \shortstack{-2.83757\\4.93732} & \shortstack{0.272538\\0} & \shortstack{-0.0105745\\0} & 0 & $4\!\times\!10^{17}$--$10^{21}$ \\
\addlinespace[2.5pt]
ARIANNA & 15 & ${10^{10}}$ & \shortstack{-1.46197\\0} & \shortstack{4.15169\\-1.41745} & \shortstack{-0.643504\\0} & \shortstack{0.0417800\\0} & 0 & $10^{16}$--$10^{20}$ \\
\addlinespace[2.5pt]
ARA & 15 & ${10^{10}}$ & \shortstack{3.93237\\0} & \shortstack{-4.24681\\8.67639} & \shortstack{0.635502\\0} & \shortstack{-0.0433609\\0} & 0 & $5\!\times\!10^{16}$--$10^{20}$ \\
\addlinespace[2.5pt]
RICE & 15 & ${10^{10}}$ & \shortstack{1.18244\\0} & \shortstack{-0.335828\\3.84467} & \shortstack{0.102555\\0} & \shortstack{-0.0102044\\0} & 0 & $10^{16}$--$10^{21}$ \\
\addlinespace[2.5pt]
FORTE & 20 & ${10^{10}}$ & \shortstack{17.1949\\0} & \shortstack{-23.0551\\38.5615} & \shortstack{2.52137\\0} & \shortstack{-0.123867\\0} & 0 & $10^{22}$--$10^{26}$ \\
\bottomrule
\end{tabular}
\end{table*}

The lunar-aperture fits require several pieces because the observing configurations cover a much larger dynamic range.  They use the same Eq.~\eqref{eq:fit-general}; the two shifted-log cases have non-zero $x_0$, and RESUN and NuMoon also require the $h[\ln(x-x_0)]^2$ term.

\begin{table*}[!htb]
\centering
\footnotesize
\setlength{\tabcolsep}{3.0pt}
\renewcommand{\arraystretch}{1.08}
\caption{Smooth-fit coefficients for the lunar-search exposures in the common form of Eq.~\eqref{eq:fit-general}.  Here $x=\log_{10}(E_\nu/{\rm eV})-q_E$, $\mathcal{E}_0$ is in ${\rm cm^2\,s\,sr}$, and the final column gives the interval over which each piece is used.  Within each coefficient column the upper and lower entries correspond to the two coefficients shown in the header.}
\label{tab:lunar-fit-coeff}
\begin{tabular}{@{}>{\raggedright\arraybackslash}p{0.225\textwidth}ccrrrrr>{\centering\arraybackslash}p{0.16\textwidth}@{}}
\toprule
Data set & $q_E$ & $\mathcal{E}_0$ & \shortstack{$a$\\$f$} & \shortstack{$b$\\$g$} & \shortstack{$c$\\$h$} & \shortstack{$d$\\$x_0$} & $e$ & $E_\nu$ range (eV) \\
\midrule
LUNASKA--ATCA & 20 & ${10^{15}}$ & \shortstack{0.604059\\0} & \shortstack{0.165257\\1.31540} & \shortstack{0.107764\\0} & \shortstack{-0.0188062\\0} & 0 & $10^{20}$--$10^{23}$ \\
\addlinespace[2.5pt]
Parkes 1996 & 20 & ${10^{15}}$ & \shortstack{2.75404\\0} & \shortstack{-5.03112\\4.14339} & \shortstack{1.73031\\0} & \shortstack{-0.216665\\0} & 0 & $10^{21}$--$10^{23}$ \\
\addlinespace[2.5pt]
Kalyazin & 20 & ${10^{15}}$ & \shortstack{-0.246811\\0} & \shortstack{-0.116531\\1.29777} & \shortstack{0.294711\\0} & \shortstack{-0.0453860\\0} & 0 & $3\!\times\!10^{20}$--$10^{23}$ \\
\addlinespace[2.5pt]
GLUE & 20 & ${10^{15}}$ & \shortstack{-0.0477078\\0} & \shortstack{0.595888\\1.00891} & \shortstack{0.0726434\\0} & \shortstack{-0.0198388\\0} & 0 & $10^{20}$--$10^{23}$ \\
\addlinespace[2.5pt]
LUNASKA--Parkes half-limb, low & 19 & ${10^{15}}$ & \shortstack{38.0697\\0} & \shortstack{-52.9415\\30.9957} & \shortstack{16.5333\\0} & \shortstack{-2.29975\\0} & 0 & $8\!\times\!10^{19}$--$10^{20.82}$ \\
\addlinespace[2.5pt]
LUNASKA--Parkes half-limb, high & 19 & ${10^{15}}$ & \shortstack{5.91219\\0} & \shortstack{-11.4641\\20.3144} & \shortstack{1.30562\\0} & \shortstack{-0.0671122\\0} & 0 & $10^{20.82}$--$10^{25}$ \\
\addlinespace[2.5pt]
LUNASKA--Parkes no-half-limb, low & 19 & ${10^{15}}$ & \shortstack{5.36274\\0} & \shortstack{-4.92434\\11.2872} & \shortstack{-2.75706\\0} & \shortstack{1.08296\\0} & 0 & $8\!\times\!10^{19}$--$10^{20.90}$ \\
\addlinespace[2.5pt]
LUNASKA--Parkes no-half-limb, high & 19 & ${10^{15}}$ & \shortstack{0.389021\\0} & \shortstack{1.07210\\1.09582} & \shortstack{-0.112760\\0} & \shortstack{0.00444604\\1.6} & 0 & $10^{20.90}$--$10^{25}$ \\
\addlinespace[2.5pt]
RESUN & 20 & ${10^{15}}$ & \shortstack{-5.33835\\0} & \shortstack{6.72424\\-2.02790} & \shortstack{-1.36225\\-0.626486} & \shortstack{0.126101\\0.8} & 0 & $7.3\!\times\!10^{20}$--$10^{24}$ \\
\addlinespace[2.5pt]
NuMoon & 20 & ${10^{20}}$ & \shortstack{-255.342\\0} & \shortstack{126.317\\-39.3490} & \shortstack{-18.3921\\-12.6412} & \shortstack{1.09816\\2.3} & 0 & $5\!\times\!10^{22}$--$10^{25}$ \\
\bottomrule
\end{tabular}
\end{table*}

As a numerical stability check, replacing a pointwise cubic interpolation of the IceCube effective area by the smooth representation of Eq.~\eqref{eq:fit-general} changes the one-decade upper limits by at most $0.33\%$.  Repeating the fixed-spectrum tests changes the quoted normalization limits by at most $0.19\%$.  These variations are much smaller than the experimental systematic uncertainties and confirm that the limits are not driven by point-to-point structure in the digitized or Monte Carlo effective-area curves.

Suppressing the common $E_\nu$ argument on the right-hand side, the pointwise sum of the published primary exposure pieces is
\begin{align}
\mathcal{E}(E_\nu)={}&
\mathcal{E}_{\rm Auger}+\mathcal{E}_{\rm IC}
+\mathcal{E}_{\rm GVD}+\mathcal{E}_{\rm KM3NeT}
+\mathcal{E}_{\rm FORTE}+\mathcal{E}_{\rm RICE}
\nonumber\\
&+\mathcal{E}_{\rm ARA}+\mathcal{E}_{\rm ARIANNA}
+\mathcal{E}_{\rm ANITA}+\mathcal{E}_{\rm RESUN}
+\mathcal{E}_{\rm NuMoon}+\mathcal{E}_{\rm GLUE}
\nonumber\\
&+\mathcal{E}_{\rm Kalyazin}+\mathcal{E}_{\rm Parkes96}
+\mathcal{E}_{\rm LUNASKA\text{-}ATCA}
+\mathcal{E}_{\rm LUNASKA\text{-}Parkes}.
\label{eq:explicit-total-exposure}
\end{align}
Here $\mathcal{E}_{\rm ANITA}$ denotes the ANITA-lite and ANITA I--IV Askaryan exposures entering the primary likelihood.  The flavor-converted ANITA-IV upward-going tau-neutrino exposure is stored separately and is added only in the two robustness variants.  

\subsection{Event counts and backgrounds}

For a one-decade test spectrum, a primary analysis contributes its published count and background only when its exposure is non-zero in the interval and the upper edge of the full decade lies within its published/reconstructed exposure range.  The non-zero inputs are summarized in Table~\ref{tab:event-backgrounds}; the remaining primary searches have $n_i=b_i=0$ for the purposes of this combination.

\begin{table}[!htb]
\centering
\caption{Non-zero event/background inputs.  The ANITA I--IV Askaryan flights are treated as one combined channel.  The upward-going ANITA-IV row is not assigned an internal diffuse-background expectation in the primary likelihood.}
\label{tab:event-backgrounds}
\begin{tabular}{lccp{0.43\linewidth}}
\hline
Analysis & $n_i$ & $b_i$ & Comment \\
\hline
KM3NeT & 1 & 0 & KM3-230213A; no background subtraction adopted here. \\
FORTE & 1 & 0 & One uncertain event; source limit uses $s_{\rm up}=3.89$ with no separate expected background mean. \\
ARA & 0 & 0.10 & Analysis A gives $0.05$ expected background events per station for A2 and A3; summed here. \\
ARIANNA & 0 & 0.08 & Approximate background per one-decade interval. \\
ANITA-IV upward-going & 4 & --- & Omitted from primary likelihood; variants use $b_i=0$ and phenomenological $b_i=4$. \\
ANITA I--IV Askaryan & 3 & 2.31 & $0+1+1+1$ events and $0+0.97+0.70+0.64$ background. \\
\hline
\end{tabular}
\end{table}

The treatment of the ANITA-IV near-horizon events is motivated in Sec.~\ref{sec:data:radio-Earth}.  In particular, $b_i=4$ for this channel is used only as a phenomenological robustness assignment and is not a collaboration-published diffuse-background estimate.

\subsection{Numerical integration and validation}

For each center $E_0$, we compute the per-channel coefficients of Eq.~\eqref{eq:alpha-bin} by integrating in $\ln E_\nu$, so that $dE_\nu/E_\nu=d\ln E_\nu$ and numerical precision is retained over the ten-decade energy range.  Before integrating, we require the high-energy edge of the full decade to lie inside the exposure range of that channel; this prevents a publication cutoff from being treated as zero acceptance over the unreported part of a bin.  The numerical integration was checked with an independent implementation.

For the quasi-differential primary limits and the ANITA-IV variant with $(n_i,b_i)=(4,0)$, the compressed likelihood is sufficiently small that the Poisson sampling distribution is summed directly rather than estimated with pseudo-experiments.  At the quoted upper limit, the resulting $p$-value agrees with 0.10 to the numerical precision of the inversion.  The five-channel ANITA-IV robustness variant with $(n_i,b_i)=(4,4)$ and the broad fixed-spectrum tests use deterministic Poisson Monte Carlo instead.  In all cases the physical precision is limited by reconstructed public effective areas/exposures and experimental systematics, so headline values are quoted to only two significant digits.  Background expectations are treated as fixed and no additional cross-experiment systematic nuisance model is imposed.

Table~\ref{tab:combined-limits} lists the primary results.  The displayed sums $N(E_0)\equiv\sum_{i\in\mathcal{P}(E_0)}n_i$ and $B(E_0)\equiv\sum_{i\in\mathcal{P}(E_0)}b_i$ are provided only for orientation: the likelihood depends on the channel-by-channel allocation of these counts and backgrounds through Eq.~\eqref{eq:product-poisson}. 

\begin{table}[!htb]
\centering
\scriptsize
\caption{Primary one-decade 90\% C.L. results.  $N$ and $B$ are the sums over active primary channels and are shown only for orientation; $\widehat Q$ and $Q_{90}$ are in $\mathrm{GeV\,cm^{-2}\,s^{-1}\,sr^{-1}}$.}
\label{tab:combined-limits}
\begin{tabular}{cccccc}
\hline
$\log_{10}(E_\nu/{\rm eV})$ interval & $E_0$ [eV] & $N(E_0)$ & $B(E_0)$ & $\widehat Q$ & $Q_{90}$ \\
\hline
16.0--17.0 & $3.162\times10^{16}$ & 1 & 0.18 & $1.649\times10^{-9}$ & $6.126\times10^{-9}$ \\
16.5--17.5 & $1.000\times10^{17}$ & 1 & 0.18 & $1.742\times10^{-9}$ & $6.471\times10^{-9}$ \\
17.0--18.0 & $3.162\times10^{17}$ & 1 & 0.18 & $1.967\times10^{-9}$ & $7.311\times10^{-9}$ \\
17.5--18.5 & $1.000\times10^{18}$ & 4 & 2.49 & $2.718\times10^{-9}$ & $9.549\times10^{-9}$ \\
18.0--19.0 & $3.162\times10^{18}$ & 4 & 2.49 & $4.564\times10^{-9}$ & $1.594\times10^{-8}$ \\
18.5--19.5 & $1.000\times10^{19}$ & 4 & 2.49 & $8.734\times10^{-9}$ & $2.958\times10^{-8}$ \\
19.0--20.0 & $3.162\times10^{19}$ & 4 & 2.49 & $1.757\times10^{-8}$ & $5.411\times10^{-8}$ \\
19.5--20.5 & $1.000\times10^{20}$ & 3 & 2.31 & $1.384\times10^{-8}$ & $1.147\times10^{-7}$ \\
20.0--21.0 & $3.162\times10^{20}$ & 3 & 2.31 & $1.608\times10^{-8}$ & $1.264\times10^{-7}$ \\
20.5--21.5 & $1.000\times10^{21}$ & 3 & 2.31 & $3.088\times10^{-8}$ & $3.661\times10^{-7}$ \\
21.0--22.0 & $3.162\times10^{21}$ & 3 & 2.31 & $0$ & $4.113\times10^{-7}$ \\
21.5--22.5 & $1.000\times10^{22}$ & 4 & 2.31 & $2.703\times10^{-7}$ & $7.565\times10^{-7}$ \\
22.0--23.0 & $3.162\times10^{22}$ & 4 & 2.31 & $1.976\times10^{-7}$ & $6.249\times10^{-7}$ \\
22.5--23.5 & $1.000\times10^{23}$ & 1 & 0.00 & $1.044\times10^{-7}$ & $4.060\times10^{-7}$ \\
23.0--24.0 & $3.162\times10^{23}$ & 1 & 0.00 & $8.188\times10^{-8}$ & $3.185\times10^{-7}$ \\
23.5--24.5 & $1.000\times10^{24}$ & 1 & 0.00 & $9.442\times10^{-8}$ & $3.673\times10^{-7}$ \\
24.0--25.0 & $3.162\times10^{24}$ & 1 & 0.00 & $1.381\times10^{-7}$ & $5.371\times10^{-7}$ \\
24.5--25.5 & $1.000\times10^{25}$ & 1 & 0.00 & $1.212\times10^{-6}$ & $4.713\times10^{-6}$ \\
25.0--26.0 & $3.162\times10^{25}$ & 1 & 0.00 & $2.190\times10^{-6}$ & $8.518\times10^{-6}$ \\
\hline
\end{tabular}
\end{table}

A numerical version of Table~\ref{tab:combined-limits} is provided as the ancillary file \texttt{limits\_A6.txt}.  Together with \texttt{total\_exposure.txt}, these are the data distributed with this article.

As a diagnostic, we also recompute an aggregate bound using the same support-valid primary channel set as the product likelihood.  Summing independent Poisson counts is mathematically valid, but doing so discards the channel identity of each event.  Around $10^{18}$--$10^{19}$~eV the channel-wise product likelihood is up to a factor of about $1.56$ more constraining than this support-matched aggregate construction because the surviving events occur in channels whose expected signal fractions differ from those of the experiments with null observations.

\subsection{Folding fixed theoretical spectra}
\label{sec:app:model-folding}

The normalization tests quoted in Sec.~\ref{sec:disc} are obtained from the same per-analysis exposures and counting inputs used for the one-decade limits, rather than by comparing a model curve point-by-point with Fig.~\ref{fig:combined-limit}.  A tabulated model is first converted to the common total all-flavor $\nu+\bar\nu$ convention.  If the source provides $E_\nu^2\,d\Phi_\nu/dE_\nu$, we interpolate $\log_{10}(E_\nu^2\,d\Phi_\nu/dE_\nu)$ as a function of $\log_{10}E_\nu$, convert back to the differential flux $\phi_{\rm model}(E_\nu)$ defined in Sec.~\ref{sec:disc}, and set it to zero outside the tabulated energy support.  We additionally impose $E_\nu\ge10^{16}$~eV and use no power-law continuation beyond the range shown by the source.  Detector exposures are folded over their published/reconstructed overlap with the model; omitted exposure beyond a detector endpoint can only increase the true signal, so the quoted fixed-spectrum normalization tests are conservative with respect to such unreported continuations.

The spectra are folded channel by channel using Eq.~\eqref{eq:model-folding} and the likelihood of Eq.~\eqref{eq:model-product}.  Because broad fixed spectra can populate a larger compressed Cartesian support, the $k_{90}$ tests are calibrated with deterministic Poisson Monte Carlo.  The two ANITA-IV robustness columns in Table~\ref{tab:model-normalizations} are recomputed after adding the upward-going exposure with either $(n_i,b_i)=(4,0)$ or the phenomenological $(n_i,b_i)=(4,4)$ assignment.

\begin{table*}[!htb]
\centering
\small
\caption{Fixed-spectrum normalization tests for the isotropic benchmark spectra considered here.  $S$ is the total nominal signal yield over the primary channels that overlap the model range.  The last two columns repeat the calculation after adding the ANITA-IV upward-going exposure with all four events as possible signal candidates [$(n_i,b_i)=(4,0)$] or with the background-tagged convention [$(n_i,b_i)=(4,4)$].  SHDM models are omitted because their Galactic contribution is anisotropic and cannot be tested with direction-averaged exposures.}
\label{tab:model-normalizations}
\begin{tabular}{lrrrr}
\hline
Model & $S(k=1)$ & primary $k_{90}$ & $k_{90}^{\tau\,{\rm cand}}$ & $k_{90}^{\tau\,{\rm bg}}$ \\
\hline
Auger cosmogenic envelope, lower & 0.0529 & 66.1 & 168 & 64.8 \\
Auger cosmogenic envelope, upper & 1.406 & 2.54 & 6.35 & 2.42 \\
TA cosmogenic best fit & 5.546 & 0.641 & 1.61 & 0.622 \\
TA cosmogenic local minimum & 2.325 & 1.49 & 3.82 & 1.46 \\
Alhebsi et al. KM3NeT-only total fit & 10.8 & 0.325 & 0.830 & 0.316 \\
Alhebsi et al. joint total fit & 2.023 & 1.74 & 4.42 & 1.71 \\
Cosmic strings, $\alpha_{\rm cs}=10^7$, $m=10^5$ GeV & 0.2612 & 24.9 & 46.5 & 24.5 \\
Cosmic strings, $\alpha_{\rm cs}=3\times10^7$, $m=4\times10^5$ GeV & 0.5556 & 12.3 & 22.7 & 12.3 \\
Cosmic strings, $\alpha_{\rm cs}=2\times10^7$, $m=10^4$ GeV & 2.023 & 3.36 & 6.16 & 3.30 \\
\hline
\end{tabular}
\end{table*}

\section{Projected sensitivities}
\label{sec:app:prospects}

This appendix documents the projected sensitivities used for comparison with the combined observational limit.  None of these curves enters the statistical combination.  The aim is to place heterogeneous projections on a common ordinate while retaining the exposure, detector configuration, and analysis level assumed by each collaboration or study.  Unless stated otherwise, the target convention is an all-flavor ($\nu+\bar\nu$) quasi-differential 90\% C.L. sensitivity per decade,
\begin{equation}
 E_\nu^2\frac{d\Phi^{90}_{\rm all}}{dE_\nu}
 = \frac{2.44\,E_\nu}{\ln 10\,\mathcal{E}_{\rm eff}(E_\nu)},
 \label{eq:prospects-common}
\end{equation}
where $\mathcal{E}_{\rm eff}$ is the exposure expressed consistently with an all-flavor flux.  The coefficient 2.44 is the Feldman--Cousins 90\% upper count for a zero-background sensitivity; below we refer to this convention as FC90.  We do not rescale different projects to a common observing time, because the assumed exposure is part of each quoted performance.

\begin{figure}[!htb]
\centering
\includegraphics[width=0.92\linewidth]{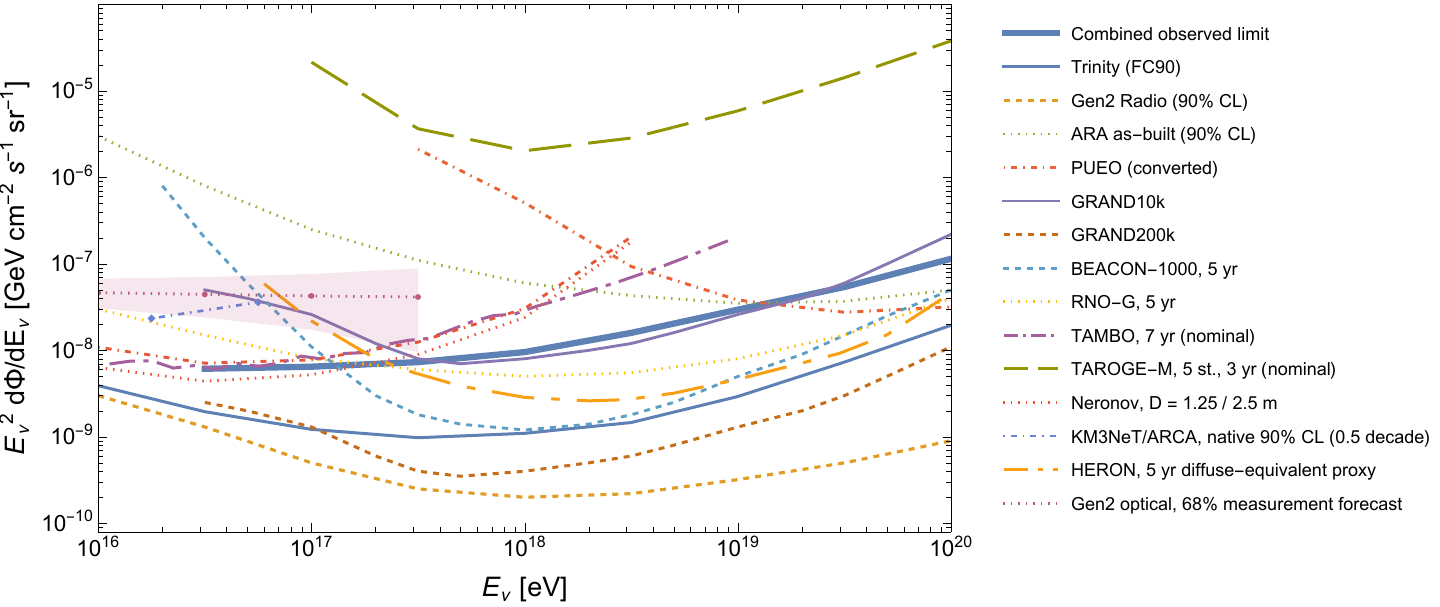}
\caption{Projected sensitivities in the $10^{16}$~eV -- $10^{20}$~eV band. See the text for details.}
\label{fig:prospects-low}
\end{figure}

\begin{figure}[!htb]
\centering
\includegraphics[width=0.92\linewidth]{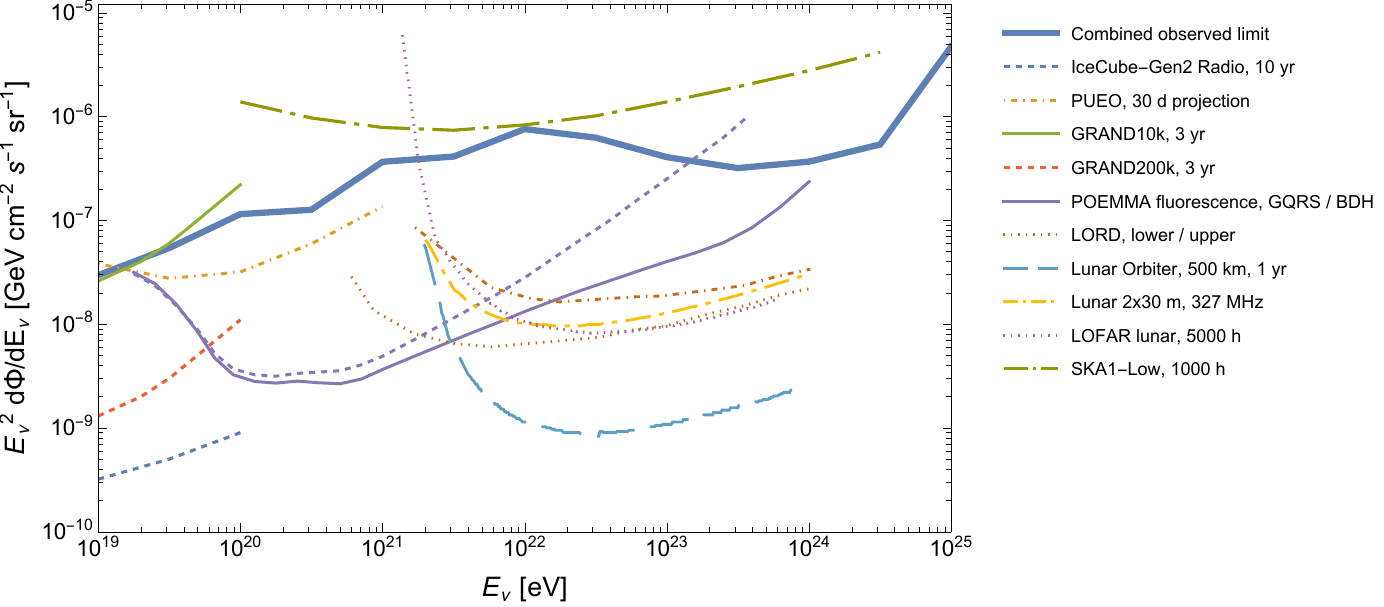}
\caption{Projected sensitivities in the $10^{19}$~eV -- $10^{25}$~eV band. See the text for details.}
\label{fig:prospects-high}
\end{figure}

The six curves retained in the main text were chosen to combine sensitivity, experimental maturity, methodological diversity, and energy coverage.  Trinity represents optical imaging of upward-going air showers from Earth-skimming $\nu_\tau$ \cite{TrinitySensitivity}; IceCube-Gen2 Radio represents a next-generation in-ice Askaryan array \cite{IceCubeGen2Radio}; PUEO represents balloon-borne radio detection of the Antarctic ice \cite{PUEOSensitivity,PUEOSimulation,PUEOWhitePaper}; GRAND10k represents a large ground-based radio array detecting very inclined air showers \cite{GRANDScience}; and SKA1-Low represents the lunar Askaryan technique at the highest energies \cite{SKALunar}.  The ARA full-array curve is included because its exposure is already realized and a diffuse analysis of these data may therefore provide a near-term improvement \cite{ARAAsBuilt,ARAArrayWideStatus2026}.  Beyond these six, the appendix comparison is intended to be more complete: Fig.~\ref{fig:prospects-low} presents the expectations for BEACON \cite{BEACONSensitivity}, RNO-G \cite{RNOGDesign}, GRAND200k \cite{GRANDScience}, KM3NeT/ARCA \cite{KM3NeTARCASensitivity}, IceCube-Gen2 Optical \cite{IceCubeGen2Optical}, HERON \cite{HERONProject}, the Neronov mountain-top concept \cite{NeronovMountain}, TAROGE-M \cite{TAROGEM}, and TAMBO \cite{TAMBO}. Figure~\ref{fig:prospects-high} retains the GRAND10k/GRAND200k, IceCube Gen2 Radio, and PUEO curves as an overlap bridge and adds POEMMA fluorescence \cite{POEMMAFluorescence} together with the additional lunar concepts LORD \cite{LORD}, Lunar Orbiter \cite{LunarOrbiter}, Lunar $2\times30$~m telescopes \cite{Lunar2x30}, and LOFAR \cite{LOFARLunar,LOFAR5000}.  Among these, KM3NeT/ARCA, IceCube-Gen2 Optical, and HERON require special treatment because their published statistical meanings are not identical to the FC90 convention of Eq.~\eqref{eq:prospects-common}.

For ARA, we use the dotted analysis-level curve in Fig.~10 of Ref.~\cite{ARAAsBuilt}.  It is a 90\% C.L. sensitivity based on the full 2013--2023 exposure and a simplified event selection, not an observed limit; the ongoing array-wide search is described in Ref.~\cite{ARAArrayWideStatus2026}.  For PUEO, we use the published 30-day trigger-level sensitivity.  The PUEO white paper defines the trigger-level single-event sensitivity with a bandwidth factor $\Delta=4$ \cite{PUEOWhitePaper}; this, together with the 90\% C.L. Feldman--Cousins count, gives the conversion factor used below.  The first PUEO flight lasted from 19 December 2025 to 12 January 2026 \cite{PUEOFlight}; we do not rescale the published 30-day curve to the realized flight duration because a final event-selection efficiency and livetime for the neutrino analysis are not yet available.

The GRAND paper provides an end-to-end simulation of the 10,000-antenna GRAND10k configuration and obtains the GRAND200k projection by scaling the direction-averaged effective area by a factor of 20 \cite{GRANDScience}.  Consequently, for the same three-year exposure and statistical convention, the corresponding GRAND10k differential sensitivity is a factor of 20 weaker than the plotted GRAND200k line.  We use the lower edge of the GRAND10k band, i.e. the aggressive-threshold scenario, for the main comparison; the larger GRAND200k extrapolation is retained only in the extended appendix plot.

Three additions in the lower-energy appendix are retained in their native statistical definitions rather than forced into FC90.  For KM3NeT/ARCA, we digitize the 10-year track-channel 90\% C.L. quasi-differential sensitivity from Ref.~\cite{KM3NeTARCASensitivity}; the published points are per flavor and use half-decade energy bins, so we apply only a factor of three for an equal-flavor all-flavor display, with no bin-width or Feldman--Cousins rescaling.  For IceCube-Gen2 Optical, we use the high-energy differential-flux forecast of Ref.~\cite{IceCubeGen2Optical}, which is a median 68\% C.L. measurement interval for independent piecewise $E_\nu^{-2}$ components, based on a combined 15-year IceCube plus 15-year Gen2 exposure.  We again apply only the factor of three from the published per-flavor ordinate and show the 68\% interval explicitly. For HERON, Ref.~\cite{HERONProject} publishes an instantaneous all-flavor \emph{fluence} sensitivity for transients rather than a diffuse upper-limit sensitivity.  We therefore digitize the native fluence curve in its published convention and, solely to indicate the relevant scale on Fig.~\ref{fig:prospects-low}, show a diffuse-equivalent proxy obtained by dividing the fluence by five years and by an instantaneous solid angle $0.06\times4\pi$~sr.  This proxy is an auxiliary conversion made here, not a HERON collaboration sensitivity and not an FC90 curve.

Tables~\ref{tab:prospects-main}, \ref{tab:prospects-extra}, and \ref{tab:prospects-extra2} summarize the normalization adopted for the curves in the extended comparison.  A factor of unity means that the published curve already uses the target convention.  ``Recalculated'' indicates that the plotted sensitivity was derived from a published aperture or exposure rather than digitized from a flux curve.  For TAROGE-M, the plot shows the nominal factor-3 all-flavor conversion listed in Table~\ref{tab:prospects-extra2}.  For TAMBO, Fig.~3 of Ref.~\cite{TAMBO} gives an all-flavor aperture consisting of the $\nu_\tau+\bar\nu_\tau$ contribution plus $\bar\nu_e$ events from the Glashow resonance.  The exact conversion to the total $\nu+\bar\nu$ all-flavor flux is therefore mildly energy dependent.  However, the TAMBO curve in Fig.~\ref{fig:prospects-low} is shown only from $10^{16}$~eV upward, above the $6.3$~PeV Glashow resonance, where the signal is dominated by tau-neutrino initiated showers.  In the plotted range, the exact all-flavor conversion is therefore practically indistinguishable from the factor-3 conversion, which is retained.

\begin{table*}[!htb]
\centering
\caption{Projected sensitivities shown in Fig.~\ref{fig:prospects-main}.  The exposures and detector configurations are kept as published.  Numerical factors refer only to conversion of the ordinate to the common all-flavor convention of Eq.~\eqref{eq:prospects-common}.}
\label{tab:prospects-main}
\small
\setlength{\tabcolsep}{3pt}
\begin{tabular}{>{\raggedright\arraybackslash}p{0.15\linewidth}>{\raggedright\arraybackslash}p{0.22\linewidth}>{\raggedright\arraybackslash}p{0.20\linewidth}>{\raggedright\arraybackslash}p{0.31\linewidth}}
\toprule
Project & Curve/configuration & Treatment & Comment \\
\midrule
IceCube-Gen2 Radio \cite{IceCubeGen2Radio} & 10 yr radio array & as published & Native 10-yr all-flavor projected sensitivity; retained at the published normalization. \\
GRAND10k \cite{GRANDScience} & 3 yr, aggressive threshold & $20\times$ GRAND200k flux sensitivity & Directly related to the simulated GRAND10k effective area used to obtain the GRAND200k extrapolation. \\
PUEO \cite{PUEOSensitivity,PUEOSimulation,PUEOWhitePaper} & 30 d trigger-level projection & $\times 4.24$ & Conversion of the published single-event/bandwidth convention ($\Delta=4$) to FC90 per decade; flight data already exist, but no diffuse-flux limit from those data is used here. \\
Trinity \cite{TrinitySensitivity} & full observatory, 10 yr & $\times 2.44$ & Native curve corresponds to one expected event; converted to FC90. \\
ARA \cite{ARAAsBuilt} & full array, 2013--2023, simplified analysis level & as published & Expected 90\% C.L. sensitivity based on realized exposure; not an observed upper limit. \\
SKA1-Low \cite{SKALunar} & lunar Askaryan, 1000 h & unit conversion and $\times 0.461$ & 1000-h SKA1-Low 90\% C.L. projection; converted from the native lunar-limit convention to the one-decade FC90 convention. \\
\bottomrule
\end{tabular}
\end{table*}

\begin{table*}[!htb]
\centering
\caption{Additional projected sensitivities retained in the extended comparison.}
\label{tab:prospects-extra}
\small
\setlength{\tabcolsep}{3pt}
\begin{tabular}{>{\raggedright\arraybackslash}p{0.15\linewidth}>{\raggedright\arraybackslash}p{0.22\linewidth}>{\raggedright\arraybackslash}p{0.20\linewidth}>{\raggedright\arraybackslash}p{0.31\linewidth}}
\toprule
Project & Curve/configuration & Treatment & Comment \\
\midrule
BEACON \cite{BEACONSensitivity} & BEACON-100/1000, 5 yr & as published & FC90 one-decade all-flavor sensitivity after the flavor conversion already included by the authors. \\
RNO-G \cite{RNOGDesign} & 35 stations, 5 yr & as published & All-flavor FC90 projection; overlaps methodologically with Gen2 Radio. \\
KM3NeT/\allowbreak ARCA \cite{KM3NeTARCASensitivity} & full ARCA, tracks, 10 yr & $\times3$ in flavor only & Native 90\% C.L. quasi-differential likelihood sensitivity in half-decade bins; no conversion to one-decade FC90. \\
IceCube-Gen2 Optical \cite{IceCubeGen2Optical} & 15 yr IceCube + 15 yr Gen2 & $\times3$ in flavor only & Median 68\% C.L.\ expected differential-flux measurement range, not an upper-limit sensitivity; the band is shown explicitly. \\
HERON \cite{HERONProject} & instantaneous transient curve; 5 yr proxy shown & native fluence retained; auxiliary $/(5\,{\rm yr}\,0.06\,4\pi)$ & Published curve is an all-flavor transient fluence sensitivity.  The plotted diffuse-equivalent proxy is derived here and should not be interpreted as FC90. \\
\bottomrule
\end{tabular}
\end{table*}

\begin{table*}[!htb]
\centering
\caption{Further prospective concepts included in the literature survey.}
\label{tab:prospects-extra2}
\small
\setlength{\tabcolsep}{3pt}
\begin{tabular}{>{\raggedright\arraybackslash}p{0.15\linewidth}>{\raggedright\arraybackslash}p{0.22\linewidth}>{\raggedright\arraybackslash}p{0.20\linewidth}>{\raggedright\arraybackslash}p{0.31\linewidth}}
\toprule
Project & Curve/configuration & Treatment & Comment \\
\midrule
POEMMA fluorescence \cite{POEMMAFluorescence} & 5 yr stereo, GQRS and BDH curves & as published & All-flavor 90\% C.L. per-decade fluorescence sensitivity; distinct from the separate POEMMA tau-viewing scenarios of Ref.~\cite{POEMMATau}. \\
GRAND200k \cite{GRANDScience} & 3 yr, full-scale projection & as published & Factor-20 effective-area extrapolation. \\
Neronov mountain-top concept \cite{NeronovMountain} & 3 yr interpretation of Fig.~5 & $\times 3_{\rm flavor}\times(2.44/3)=\times 2.44$ & Fig.~5 is interpreted as $N=1$ in one year, hence $N=3$ in three years; this is a power-law envelope, not a bin-by-bin differential sensitivity. \\
LORD \cite{LORD} & upper/lower scenarios & $\times 0.461$ & Converts the native $\Delta\ln E_\nu=1$ convention with 2.3 events to FC90 per decade. \\
Lunar Orbiter \cite{LunarOrbiter} & 500 km, 1 yr & $\times 0.461$ & Same logarithmic-bin conversion as LORD. \\
Lunar $2\times30$ m \cite{Lunar2x30} & 327 MHz, 4000 h & $\times 0.461$ & Same logarithmic-bin conversion as LORD. \\
LOFAR \cite{LOFARLunar,LOFAR5000} & 150 MHz; 5000 h adopted & recalculated from aperture & The aperture paper specifies no unique survey duration; 5000 h follows the later phenomenological forecast. \\
TAROGE-M \cite{TAROGEM} & 5 stations, 3 yr & recalculated from $\nu_\tau$ exposure & The plotted appendix curve uses the nominal factor-3 all-flavor conversion. \\
TAMBO \cite{TAMBO} & nominal 7 yr & recalculated from aperture & Fig.~3 includes $\nu_\tau+\bar\nu_\tau$ plus the $\bar\nu_e$ Glashow contribution.  Since the plotted range starts at 10~PeV, above the 6.3~PeV resonance, the factor-3 all-flavor conversion is retained. \\
\bottomrule
\end{tabular}
\end{table*}

The extended comparison is split into two overlapping energy intervals to avoid obscuring the physically relevant parts of the curves.  Consistent with the scope of this work, the lower panel starts at $10^{16}$~eV and extends to $10^{20}$~eV, concentrating on the  region where the optical, in-ice, balloon, and ground-radio concepts have their greatest leverage.  The higher panel spans $10^{19}$--$10^{25}$~eV and emphasizes POEMMA and the highest-energy radio and lunar techniques.  The one-decade overlap is intentional: Gen2 Radio, PUEO, GRAND10k, and GRAND200k provide a direct visual bridge between the two panels.  Together the two panels are intended to include all projects from Tables~\ref{tab:prospects-main}, \ref{tab:prospects-extra}, and \ref{tab:prospects-extra2}, for which a quantitative projected sensitivity or a defensible proxy could be recovered in machine-readable form.

\clearpage
\bibliographystyle{nsr}
\bibliography{vhe}

\end{document}